\documentclass[12pt,a4paper]{article}
\usepackage{ucs}
\usepackage[utf8x]{inputenc}
\usepackage[english]{babel}
\usepackage[T1]{fontenc}
\usepackage{ae,aecompl}
\usepackage[numbers,comma,square,compress]{natbib}
\usepackage{amsmath,amssymb,amsfonts}
\usepackage{ifpdf}
\ifpdf
  \usepackage[pdftex,a4paper,hmargin=25mm,vmargin=24mm]{geometry}
  \usepackage[pdftex]{hyperref}
\else
  \usepackage[dvips,a4paper,hmargin=25mm,vmargin=24mm]{geometry}
  \usepackage[dvips]{hyperref}
\fi

\newcommand{\email}[1]{\href{mailto:#1}{#1}}

\newcommand{\be}{\begin{equation}}
\newcommand{\ee}{\end{equation}}
\newcommand{\bea}{\begin{eqnarray}}
\newcommand{\eea}{\end{eqnarray}}
\newcommand{\nn}{\nonumber}

\newcommand{\mr}[1]{\mathrm{#1}}

\newcommand{\cS}{\mathcal{S}}
\newcommand{\cN}{\mathcal{N}}

\begin{document}
\begin{center}
{\Large On entropic gravity: the entropy postulate, entropy content\\
\vspace{.3em}
 of screens and relation to quantum mechanics}\\
\vspace{1.5em}
\textbf{M. Chaichian\footnote{\email{masud.chaichian@helsinki.fi}}, 
M. Oksanen\footnote{\email{markku.oksanen@helsinki.fi}} and 
A. Tureanu\footnote{\email{anca.tureanu@helsinki.fi}}}\\
\vspace{1em}
\textit{Department of Physics, University of Helsinki,\\
 P.O. Box 64, 00014 Helsinki, Finland}
\end{center}

\vspace{1em}
\begin{abstract}
We consider the controversial hypothesis that gravity is an entropic force that has its origin in the thermodynamics of holographic screens. Several key aspects of entropic gravity are discussed. In particular, we revisit and elaborate on our criticism of the recent claim that entropic gravity fails to explain observations involving gravitationally-bound quantum states of neutrons in the GRANIT experiment and gravitationally induced quantum interference. We argue that the analysis leading to this claim is troubled by a misinterpretation concerning the relation between the microstates of a holographic screen and the state of a particle in the emergent space, engendering inconsistencies. A point of view that could resolve the inconsistencies is presented. We expound the general idea of the aforementioned critical analysis of entropic gravity in such a consistent setting. This enables us to clarify the problem and to identify a premise whose validity will decide the faith of the criticism against entropic gravity. It is argued that in order to reach a sensible conclusion we need more detailed knowledge on entropic gravity. These arguments are relevant to any theory of emergent space, where the entropy of the microscopic system depends on the distribution of matter in the emergent space.
\end{abstract}

\vspace{.5em}
\begin{tabular}{ll}
\textbf{Keywords}: &entropic gravity, entropic force, emergent gravity, emergent space,\\
&quantum mechanics
\end{tabular}
\vspace{.5em}

\section{Introduction}
We consider the entropic gravity (EG) hypothesis proposed by Verlinde \cite{Verlinde}, where gravity is an emergent phenomenon driven by the second law of thermodynamics: entropy increases until a thermodynamic equilibrium is reached. In EG, space, inertia and gravity are postulated to emerge from the thermodynamics of an unknown microscopic theory of holographic screens.
This proposal was preceded by a considerable amount of research on the relation between gravity and thermodynamics, and by various attempts to give gravity a thermodynamic reinterpretation. These studies have been heavily motivated by the advent of black hole thermodynamics \cite{Bekenstein,Bardeen,Hawking,Davies,Unruh}. The idea of holography was originally introduced in Refs. \cite{tHooft,Susskind}. In the seminal paper \cite{Jacobson}, the Einstein equation was derived locally on Rindler causal horizons as a thermodynamic equation of state (see also  Refs. \cite{Jacobson2}). Other major contributions to the study of holographic and thermodynamic aspects of gravity were made in Refs. \cite{Padmanabhan,Padmanabhan2}. Among other things, the (holographic) relation of bulk and surface terms in gravitational actions were extensively studied in these latter works, arguing that the field equations of any diffeomorphism invariant theory of gravity have a thermodynamic reinterpretation and showing that the equipartition of energy in the microscopic degrees of freedom of a Rindler horizon can be used to derive gravity. These results suggest that gravity and spacetime might be emergent concepts which may have a thermodynamic origin.
The new insight of Ref. \cite{Verlinde} is to recognize that the entropy of a holographic screen can change due to the displacement of matter that is located far away from the screen. When a particle moves closer to a screen, the entropy density of the screen increases. In the presence of a nonzero temperature on the screen, this leads to an attractive entropic force that can be identified as gravity. Thus, in addition to storing the information that describes the world inside a screen, holographic screens also have to contain some information about the world outside. Clearly, the EG hypothesis is still heuristic at the moment.

In Ref. \cite{ourEGnote}, we analyzed critically the treatment of neutron states in Ref. \cite{KobakhidzeEG}, where it was argued that EG fails to explain the observation of gravitationally-bound quantum states of neutrons. Extremely fine observations of the two lowest energy states of neutrons in a quantum bouncer formed by the Earth's gravitational field and a neutron mirror were performed in the GRANIT experiment \cite{N,N2} (for further analysis of the experiment, see also Refs.~\cite{Voronin,Westphal,N3,Pignol}). A method for observing magnetically-induced resonance transitions between gravitationally-bound quantum states of neutrons in the GRANIT spectrometer has been presented in Ref.~\cite{Kreuz}, which could provide a way to measure the higher energy levels. An experiment that realizes resonance transitions between the gravitationally-bound neutron states by introducing a mechanically vibrating neutron mirror has been reported recently \cite{Jenke}. We concluded that EG does not necessarily contradict the results of the GRANIT experiment, since it is conceivable that the holographic description assumed in EG could produce not only gravity but quantum mechanics as well. Indeed the idea of holography is that everything inside a screen is an image of the data that is stored and processed on the screen.

In this Letter, we elaborate on the point of view of our criticism \cite{ourEGnote} and discuss the recent communication \cite{KobakhidzeEG2}, where the conclusion of \cite{KobakhidzeEG} is restated and it is also argued that the coherence and interference of quantum states is destroyed in EG, so that EG not only fails to explain the results of the GRANIT experiment but also, for example, the gravitationally-induced quantum interference \cite{Colella}.\footnote{This claim was already present in Ref.~\cite{KobakhidzeEG} implicitly, where it is argued that the coherence of any state that extends in the direction of gravity is destroyed in EG.} The result of Refs.~\cite{KobakhidzeEG,KobakhidzeEG2} is based on the argument that the size of the state space that describes a particle necessarily changes with the distance to another particle.
We argue that a premise of the analysis of Refs.~\cite{KobakhidzeEG,KobakhidzeEG2} is based on a misinterpretation concerning the relation of the microstates of a holographic screen and the state of a particle in the emergent space. This premise is the assumption that the state of a particle at position $\vec{r}$ is described by the density operator that consists of fragments of the microstates of the holographic screen that includes the point $\vec{r}$. This assumption leads to at least two inconsistencies, which we have briefly pointed out in Ref.~\cite{ourEGnote}.
We hope this observation will help us find a consistent way to accomplish such an important analysis.

First we discuss the interpretation of the fundamental entropy postulate of EG in Sect.~\ref{sec.2} and the entropy content of screens in Sect.~\ref{sec.3}. In these two sections we expose the two inconsistencies that are implied by the aforementioned assumption concerning the description of the state of a particle in Refs. \cite{KobakhidzeEG,KobakhidzeEG2}. In Sect.~\ref{sec.4} we elaborate on the arguments of our paper \cite{ourEGnote} and confirm that the description of the state of a particle used in Refs. \cite{KobakhidzeEG,KobakhidzeEG2} indeed leads to two inconsistencies. We also give detailed answers to the counterarguments presented in Ref. \cite{KobakhidzeEG2}. In Sect.~\ref{sec.5} we discuss a point of view where the inconsistencies concerning the state of a particle can be avoided. The general idea of the argument in Refs.~\cite{KobakhidzeEG,KobakhidzeEG2}, and the decisive premise (see \eqref{fragmentpremise}) behind it, are expounded and discussed. We also consider the meaning of this argument in a generic theory of emergent space. Sect.~\ref{sec.6} contains the conclusions.

We note that in addition to the papers \cite{KobakhidzeEG,KobakhidzeEG2}, EG has also been criticized by other authors. Arguments against EG together with some
clarifying comments and plausible ways out have been presented, e.g., in Refs. \cite{Hossenfelder,Gao,Li,Hu,Visser} and \cite{ourEGnote}.

\section{Interpretation of the entropy postulate}\label{sec.2}
First we briefly review the EG hypothesis \cite{Verlinde}. Consider the fundamental entropy postulate of EG \cite{Verlinde}:
\be\label{DeltaS}
\Delta S=2\pi m \Delta r \,.
\ee
We assume units in which $\hbar=c=k_B=1$.
What does the formula \eqref{DeltaS} mean? In \eqref{DeltaS}, $\Delta S$ is the increase in the entropy of a holographic screen $\cS$ when a test particle, which has the mass $m$ and is located at the distance $\Delta r$ from $\cS$, moves to the immediate vicinity of $\cS$. It is assumed that the information stored on the screen is somehow affected due to the approaching particle, so that its entropy changes according to \eqref{DeltaS}. Finally the particle $m$ merges into $\cS$, essentially becoming a part of the information and energy on the screen.
This interpretation of the entropy postulate \eqref{DeltaS} has some similarities with Bekenstein's famous thought experiment on black hole entropy \cite{Bekenstein}. Because of the increase in entropy \eqref{DeltaS} associated with the displacement $\Delta r$ of the particle towards $\cS$, there is a statistical tendency for the particle to be closer to $\cS$. This leads to an attractive entropic force $F$ that is defined by
\be\label{F.Deltar}
F\Delta r = T\Delta S \,,
\ee
where $T$ is the temperature of $\cS$. This entropic force can be identified as gravity. From \eqref{DeltaS} and \eqref{F.Deltar} we see that the gravitational acceleration is defined by the temperature of $\cS$ as $g=2\pi T$. In other words, $T$ is equal to the Unruh temperature $T=g/2\pi$. More generally, the particle $m$ can be located at any distance from the screen. Then an infinitesimal displacement $\delta\vec{r}$ of the particle $m$ is associated with a change $\delta dS$ in the entropy density of the screen $\cS$, and the resulting entropic force $\vec{F}$ is defined by \cite{Verlinde,Hossenfelder}
\be
\vec{F}\cdot\delta\vec{r} = \int_\cS T \delta dS \,,
\ee
where the integral is taken over a screen that does not contain the particle at $\vec{r}$.

The number of microscopic degrees of freedom $N$ on a screen is proportional to the area $A$ of the screen as
\be\label{N}
N=\frac{A}{G} \,,
\ee
where $G$ is the gravitational constant, i.e. Planck length squared. The average energy of a microscopic degree of freedom is assumed to be defined by the temperature on the screen according to the equipartition rule\footnote{It has been pointed out that the equipartition rule needs to be corrected at very low temperatures due to the quantization of the energy of the microscopic degrees of freedom \cite{CGao,Kiselev}. It has also been argued that Newton's gravitational law is dramatically altered for high gravitational fields if the energy of the microscopic degrees of freedom is bounded \cite{Sahlmann}.} $\langle E_\mr{d.o.f.}\rangle=\frac{1}{2}T$.
Hence, the total energy $E$ of the screen, which is equal to the mass $M$ it contains, is given by
\be\label{M}
M=E=\frac{1}{2}NT \,,
\ee
assuming the energy is evenly distributed over the microscopic degrees of freedom.
In the more general description one writes
\be
\int_{(\cS)}\varrho dV=\frac{1}{2}\int_\cS TdN=\frac{1}{2G}\int_\cS TdA \,,
\ee
where $\varrho$ is the mass density and $\int_{(\cS)}dV$ denotes the integral over the volume enclosed by $\cS$.

Consider a system that consists of a particle of mass $M$ at the origin and a spherical holographic screen $\cS_r$ of radius $r$ around $M$ and a test particle of mass $m$ at $r+\Delta r$. The temperature of the screen $\cS_r$ can be obtained from \eqref{N} and \eqref{M} as:
\be\label{NMT}
T=\frac{2M}{N(r)}=\frac{GM}{2\pi r^{2}} \,,
\ee
where the number of microscopic degrees of freedom on $\cS_r$ is
\be\label{N(r)}
N(r)=\frac{4\pi r^{2}}{G} \,.
\ee

Now let us compare our understanding of the entropy postulate \eqref{DeltaS} to the interpretation of Refs. \cite{KobakhidzeEG,KobakhidzeEG2}. In Ref. \cite{KobakhidzeEG2}, it is stated that the test particle $m$ at position $r$ is described by a statistically large number $n(r)$ of microstates, which depends on the position of the particle with respect to the other particle $M$:
\be\label{n(r)}
n(r)=\frac{2m}{T}=\frac{4\pi r^{2}}{G}\frac{m}{M}=\frac{m}{M}N(r) \,.
\ee
Hence the test particle $m$ carries the entropy $S_m(r)$ which changes with the distance $r$ as:\footnote{In this Letter we denote the particle $m$ with the subscript $m$ instead of $\cN$, because here it is not necessary to specifically consider a neutron.}
\be\label{DeltaSm}
\Delta S_m=\Delta \log n(r)=\Delta \log N(r)=\Delta S=2\pi m \Delta r \,.
\ee
In summary, it is claimed that the number of microstates that constitute the state space of the particle $m$ grows with the distance to the particle $M$. Therefore the entropy of the state of the particle $m$ increases with the distance as well. Specifically the entropy $S_m$ of the particle $m$ is considered to increase as in Eq.~\eqref{DeltaSm} when the distance to the particle $M$ is increased by $\Delta r$.

We argue that these two interpretations of the entropy postulate of EG are not consistent with each other. In Eq.~\eqref{DeltaS}, $\Delta S$ is the increase in the entropy $S_\cS$ of the screen $\cS$, $\Delta S_\cS=\Delta S$, when the test particle $m$ is displaced closer to the screen. Nothing is said about the entropy of the particle $m$. Indeed, it is assumed that the entropy of the particle $m$ does not change upon the displacement. When the particle $m$ merges into $\cS$ we should, however, take into account the entropy of the particle. If the particle $m$ has entropy $S_m$, the entropy of the screen $\cS$ has to increase by the amount $\Delta S_\cS=\Delta S+S_m$ when the particle merges into $\cS$, so that the change of the total entropy is $\Delta S_\mr{total}=(S_\cS+\Delta S_\cS)-(S_\cS+S_m)=\Delta S$ and consequently the entropic force remains the same as before. In other words when the particle $m$ merges into $\cS$, its entropy $S_m$ is merged into $\cS$ as well, and in addition there is the increase \eqref{DeltaS} in $S_\cS$ that is associated with the displacement of the particle $m$: $r+\Delta r\rightarrow r$. This increase in total entropy again means that there exists an entropic force, $F\Delta r = T\Delta S_\mr{total} = T\Delta S$.

In Refs. \cite{KobakhidzeEG,KobakhidzeEG2} the interpretation is very different, because the entropy difference \eqref{DeltaS} is associated with the state of the test particle $m$. The entropy $S_m(r+\Delta r)$ of the particle $m$ increases with the distance $\Delta r$ to the screen $\cS_r$ as given in \eqref{DeltaSm}. Let us consider what happens to the entropic force in this interpretation. The entropy of the screen is $S_{\cS_r}$. Since the total entropy $S_\mr{total}(r,\Delta r)=S_{\cS_r}+S_m(r+\Delta r)$ increases monotonically with the distance $\Delta r$ as $\Delta S_\mr{total}=\Delta S_m=\Delta S=2\pi m \Delta r$, there is a statistical tendency for the test particle to move away from the screen, i.e. the entropic force \eqref{F.Deltar} is repulsive. This clearly contradicts the idea of EG.
One could obtain an attractive entropic force by considering that the entropy of the particle decreases monotonically with the distance to the screen. But in that case the entropy of the particle could become negative provided that the initial entropy of the particle was sufficiently low, which makes no sense.

Now we know that the entropy of the test particle $m$ cannot depend on  its distance to the screen in the way it is claimed in Eqs. \eqref{n(r)} and \eqref{DeltaSm}. But let us still check the calculation of \eqref{DeltaSm}. We obtain from \eqref{n(r)}:
\be\label{DeltaLogN}
\Delta \log n(r)=\log\frac{n(r+\Delta r)}{n(r)}=2\log\left( 1+\frac{\Delta r}{r} \right)\neq 2\pi m \Delta r=\Delta S \,.
\ee
Thus there is something wrong in the calculation of \eqref{DeltaSm}. Obviously \eqref{DeltaLogN} does not change if we replace $n(r)$ with $N(r)$ \eqref{N(r)}.
However, this problem \eqref{DeltaLogN} with \eqref{DeltaSm} is not essential to our arguments. The essential point is that according to Refs.~\cite{KobakhidzeEG,KobakhidzeEG2} the entropy difference \eqref{DeltaS} is associated with the state of the particle $m$. It is claimed to have $\Delta S_m=2\pi m\Delta r$ more entropy at $r+\Delta r$ than at $r$.

Then let us discuss the meaning of $n(r)$ in \eqref{n(r)} and the validity of its use in Ref. \cite{KobakhidzeEG2}. $n(r)$ is the (approximate) number of the microscopic degrees of freedom that describe the particle $m$ on the screen $\cS_r$, when the particle is contained within the screen. For simplicity, it has been assumed $m\lll M$, so that the particle $m$ does not affect the geometric form and temperature of the screen significantly.
Note that $n(r)$ is defined by the radius $r$ of the screen $\cS_r$, not by the position of the particle. The particle could be anywhere inside the screen. The equation \eqref{n(r)} for $n(r)$ only tells us that a given piece of matter is described by many more microscopic degrees of freedom on a larger holographic screen than it is on a smaller screen.
We do not yet know how the microscopic degrees of freedom on a holographic screen are related to the state of the particle in the emergent space. In particular, the fact that a test particle $m$ at $r$ is described by $n(r)$ microscopic degrees of freedom on the screen $\cS_r$ does not mean that the density operator of the particle at $r$ consists of $n(r)$ states. It is important to remember that the number of microscopic degrees of freedom and the number of microstates are different things. These points will be further discussed in the following sections. Thus, although the equation \eqref{n(r)} is correct in the sense we have explained above, it is misinterpreted in Ref. \cite{KobakhidzeEG2}.

\section{The entropy content of screens}\label{sec.3}
We wish to repeat our concern about the claimed entropy content of holographic screens given in Refs. \cite{KobakhidzeEG,KobakhidzeEG2}. First it is stated that a holographic screen contains maximum entropy that can be fitted in the volume surrounded by the screen. Later it is assumed that the entropy of a screen is determined by the physical system which the screen contains and describes. These two statements cannot hold at the same time. The first statement is clearly false, because only black hole horizons can be considered to have maximal entropy. The latter assumption is not fully accurate either.

Let us first explain what we mean by the entropy of a holographic screen. A microstate of a holographic screen consists of the states of all the microscopic degrees of freedom on the screen. Each microstate of a holographic screen corresponds to a mass distribution in actual space, which is obtained by coarse graining the microstate as far as possible (until the screen or each compact part of the screen reduces to minimal area or to a horizon of a black hole, whichever is discovered first). The macrostate of the screen consists of all the microstates that reduce to the given mass distribution within the screen via coarse graining. The macrostate is represented by a density operator that consists of all those microstates that belong to the macrostate. The entropy of a macrostate of the screen is defined as the logarithm of the number of microstates in the macrostate. In other words the entropy of a screen is the maximal entropy that can be associated with the matter distribution within the screen.

Thus the macrostate and entropy of a screen is determined by the mass distribution that is contained within the screen. However, two screens that contain the same mass configuration, but different amounts of empty space around the mass, have different areas. Therefore the screens can have different amounts of microstates associated with their macrostates. This means that the screens can have different amounts of entropy. Thus, in the description of entropy we just gave, ``empty space'' is an essential part of a ``mass distribution'' that determines the macrostate of a screen, and consequently its entropy. As an example consider the spherical screen $\cS_r$ whose macrostate corresponds to a given spherically symmetric mass distribution of total mass $M$ at the center of the screen. The screen has $N(r)$ microscopic degrees of freedom and its macrostate consists of $\Omega_{\cS_r}$ microstates. The number of microscopic degrees of freedom grows monotonically with the radius $r$ of the screen. Consequently the number of available microstates $\Omega_{\cS_r}$ can grow with the radius $r$ of the screen too. We can quantify the rate of the growth of the entropy of the screens,  $S_{\cS_r}=\log\Omega_{\cS_r}$, by recalling that the entropy per degree of freedom decreases according to the coarse graining variable $-2\Phi(r) = 2GM/r$ \cite{Verlinde} and using the known number of microscopic degrees of freedom \eqref{N(r)}. Up to a constant we obtain $S_{\cS_r} = -\frac{1}{2}\Phi(r)N(r) = 2\pi Mr$, where the normalization is chosen to match the entropy of a black hole horizon. The number of microstates in the macrostate of the screen increases exponentially with the radius of the screen, $\Omega_{\cS_r}=b^{2\pi Mr}$, where $b$ is the base of the logarithm. Thus, a larger holographic screen around the mass $M$ has greater entropy than a smaller screen.

\section{The problems with the state of a test particle}\label{sec.4}
In this section we elaborate some of the critical arguments, which we presented in Ref.~\cite{ourEGnote}, concerning the state of a test particle in Ref. \cite{KobakhidzeEG}. We also provide answers to the rebuttal of our arguments found in Ref. \cite{KobakhidzeEG2}.

We use the following notation. The screen at $r+\Delta r$ which also contains the particle $m$ at $r+\Delta r$ is denoted by $\cS^{\supset m}_{r+\Delta r}$ and its density operator by $\rho_{\cS^{\supset m}_{r+\Delta r}}$. The density operator of the screen $\cS_r$ at $r$ is denoted by $\rho_{\cS_r}$. The density operator of the fragment of $\cS^{\supset m}_{r+\Delta r}$ which is supposed to describe the particle $m$ at $r+\Delta r$ is denoted by $\rho_m(r+\Delta r)$. The fragment $\rho_m(r+\Delta r)$ consists of the states of the $n(r+\Delta r)$ microscopic degrees of freedom that are supposed to describe the particle $m$ on $\cS^{\supset m}_{r+\Delta r}$. One can say that $\rho_m(r+\Delta r)$ consists of fragments of all the microstates in $\rho_{\cS^{\supset m}_{r+\Delta r}}$.

First we have to emphasize that we are not convinced that the state of the particle $m$ in the emergent space at $r+\Delta r$ can be consistently described by the fragment $\rho_m(r+\Delta r)$ of the screen $\cS^{\supset m}_{r+\Delta r}$. It is against the idea of holography to directly associate such microscopic degrees of freedom on a screen with the degrees of freedom of a particle in actual space. Moreover, it is unclear how one is supposed to separate the microscopic degrees of freedom that describe the particle $m$ from all the degrees of freedom on the screen. After all, we have no knowledge on the microscopic theory and the microscopic physics is supposed to look random. But for the sake of argument, let us assume that $\rho_m(r+\Delta r)$ does describe the state of the particle $m$ at $r+\Delta r$, and see where it leads.

In Ref.~\cite{ourEGnote}, we did not argue that \eqref{DeltaSm} implies that holographic screens at different $r$ have the same number of microstates, and hence their entropy does not depend on $r$. In fact, we argued that the assumption of Ref. \cite{KobakhidzeEG} (also repeated in \cite{KobakhidzeEG2}) that the states $\rho_m(r+\Delta r)\otimes\rho_{\cS_r}$ and $\rho_{\cS^{\supset m}_{r+\Delta r}}$ are equal -- in the sense that they describe the same physical state and have the same entropy -- implies the following two things:
\begin{enumerate}
\item\label{implication:1} The entropy difference \eqref{DeltaS} would be solely associated with the fragment $\rho_m(r+\Delta r)$ that describes the particle $m$. This would mean that the entropy of the particle $m$ depends on its distance $\Delta r$ to the screen. Apparently Ref. \cite{KobakhidzeEG2} agrees with this conclusion by stating that the entropy \eqref{DeltaS} is carried by the particle $m$, as we cite in and above Eq. \eqref{DeltaSm}.
\item\label{implication:2} Screens $\cS_r$ with different radii $r$ would have the same state, $\rho_{\cS_r}=\rho_{\cS_{r+\Delta r}}$, and hence the same entropy, $S_{\cS_r}=S_{\cS_{r+\Delta r}}$. The removal of the particle $m$ from the states $\rho_m(r+\Delta r)\otimes\rho_{\cS_r}$ and $\rho_{\cS^{\supset m}_{r+\Delta r}}$ indeed gives the states $\rho_{\cS_r}$ and $\rho_{\cS_{r+\Delta r}}$, respectively, as will be explained shortly.
\end{enumerate}
We have shown that both of these implications are inconsistent with EG: see Sections \ref{sec.2} and \ref{sec.3} for the implications \ref{implication:1}. and \ref{implication:2}., respectively. Therefore the two states $\rho_m(r+\Delta r)\otimes\rho_{\cS_r}$ and $\rho_{\cS^{\supset m}_{r+\Delta r}}$ cannot be the same physical state and their entropies need not be equal.

It was claimed \cite{KobakhidzeEG2} that removing the particle $m$ from the density operator $\rho_{\cS^{\supset m}_{r+\Delta r}}$ of the screen $\cS^{\supset m}_{r+\Delta r}$ corresponds to coarse graining that reduces the density operator to the one $\rho_{\cS_r}$ of the screen $\cS_r$. However, the coarse graining of $\rho_{\cS^{\supset m}_{r+\Delta r}}$ to $\rho_{\cS_r}$ also coarse grains the microscopic degrees of freedom which describe the mass $M$, not only the microscopic degrees of freedom which describe the particle $m$. After all, most of the microscopic degrees of freedom on the screen describe the mass $M$ and their number decreases with the radius of the screen as in \eqref{N(r)}. Therefore, contrary to what is claimed in Ref. \cite{KobakhidzeEG2}, the removal of the particle $m$ is not the same as the coarse graining of $\rho_{\cS^{\supset m}_{r+\Delta r}}$ to $\rho_{\cS_r}$.

Let us consider what happens when we remove the particle $m$ from the states $\rho_m(r+\Delta r)\otimes\rho_{\cS_r}$ and $\rho_{\cS^{\supset m}_{r+\Delta r}}$. We agree with Ref. \cite{KobakhidzeEG2} that removing the particle $m$ from the state $\rho_m(r+\Delta r)\otimes\rho_{\cS_r}$ gives the state $\rho_{\cS_r}$. What happens to the state $\rho_{\cS^{\supset m}_{r+\Delta r}}$ when we remove the particle $m$? For this case it is more appropriate and illustrative to talk about bits rather than microscopic degrees of freedom. Imagine the screen is a device that stores and processes information. Removing the particle $m$ means that we delete the bits that describe the particle $m$ on the screen $\rho_{\cS^{\supset m}_{r+\Delta r}}$. After those bits are deleted, the freed storage space is free to describe the remaining physical content of the screen, namely the mass $M$, along with all the rest bits on the screen that already described $M$. The energy of the screen decreases by the small amount $m$ and consequently the temperature of the screen decreases correspondingly \eqref{NMT}. The amount of storage space on the screen does not change, because it is defined by the area of the screen, which remains constant. Thus the result is the state $\rho_{\cS_{r+\Delta r}}$ of the screen $\cS_{r+\Delta r}$, where the particle $m$ is no longer present in the system. This confirms the ``implication \ref{implication:2}.''.

We should make a few remarks about the coarse graining of holographic screens. When the screen $\cS^{\supset m}_{r+\Delta r}$ is coarse grained (by the amount $\Delta r$) it reduces to the screen $\cS_r$ and at the same time the spherical shell of space at $]r,r+\Delta r]$ emerges together with the particle $m$ at $r+\Delta r$, or more generally with all the matter that resides in the emerged part of space. Thus when the state $\rho_{\cS^{\supset m}_{r+\Delta r}}$ is coarse grained, the resulting state does not only include the state $\rho_{\cS_r}$ but also the state of the particle $m$ in the emerged space at $r+\Delta r$.  As we can see the coarse graining does not remove the particle $m$ from the system, but rather transforms it from information living in the screen $\cS^{\supset m}_{r+\Delta r}$ to a particle living in the actual space. The state of the particle in the emergent space is not related to the microscopic degrees of freedom of screens in a trivial way.
Finally, we wish to remark that it would be highly questionable to talk about the coarse graining of the state $\rho_m(r+\Delta r)\otimes\rho_{\cS_r}$, because this state is not a state of a screen, but instead a product of a screen $\cS_r$ and a fragment of another screen $\cS^{\supset m}_{r+\Delta r}$ that encloses the screen $\cS_r$. It is unclear whether the coarse graining procedure -- whatever it is -- can be applied to such composite states that include screens and their parts which reside within each other.

In order to see that $\rho_m(r+\Delta r)\otimes\rho_{\cS_r}$ and $\rho_{\cS^{\supset m}_{r+\Delta r}}$ are not the same physical state, it is not necessary to consider the removal of the particle $m$.  Alternatively we can convince ourselves that these states have different amounts of entropy, which shows that the states cannot be equal. Compare the amount of information that these two states reveal to an outside observer. It is clear that the state $\rho_m(r+\Delta r)\otimes\rho_{\cS_r}$ reveals more information on the position of the particle $M$ (or more generally on the positional distribution of the mass $M$) to an outside observer than the state $\rho_{\cS^{\supset m}_{r+\Delta r}}$, simply because the screen $\rho_{\cS_r}$ confines the mass $M$ to a smaller space than the screen $\cS^{\supset m}_{r+\Delta r}$. According to the fundamental relation between available information and entropy, more available information means less entropy. Thus, assuming the states $\rho_m(r+\Delta r)\otimes\rho_{\cS_r}$ and $\rho_{\cS^{\supset m}_{r+\Delta r}}$ reveal a similar amount of information on the particle $m$, the state $\rho_{\cS^{\supset m}_{r+\Delta r}}$ must have more entropy than the state $\rho_m(r+\Delta r)\otimes\rho_{\cS_r}$ has.
This is an expected result, because on a bigger screen one can have many more microstates which correspond to a given macrostate than on a smaller screen (recall the discussion in Sect.~\ref{sec.3}).

\section{Is a consistent description of quantum particles possible in entropic gravity?}\label{sec.5}
In the previous sections we have discussed the inconsistencies that arise if we assume that the state of a particle at position $\vec{r}$ is described by a fragment of the holographic screen that includes the point $\vec{r}$. Indeed, there is no compelling reason why the particle at $\vec{r}$ should be exclusively described by the screen that includes the point $\vec{r}$. The particle can be described on any screen that contains the particle. Since the inconsistencies are related to the use of several holographic screens in the description of the state of the test particle, we will avoid them by describing the system with a single holographic screen. This way we will clarify the problem whether it is possible to describe a quantum particle consistently in entropic gravity.

When we consider a classical particle, we can always tell whether a given screen contains the particle. For a quantum particle that is no longer possible for all screens, since the position of the particle is not defined precisely. This gives us a reason to argue that the screen that includes the point $\vec{r}$ may not contain all the necessary information on the state of a particle at $\vec{r}$. Therefore, in order to describe the state of the particle, it makes more sense to choose a screen that is large enough to contain the particle with certainty (or at least with very high probability). In other words to choose a screen that contains the hole system that is relevant to the state of the particle. Let us again consider the system that consists of the fixed mass $M$ at origin and the test particle $m$ at $\vec{r}$. We assume the particle $m$ is confined between the screen $\cS_{r_M}$ at $r_M$ and the screen $\cS^{\supset m}_{r_\mr{abs}}$ at $r_\mr{abs}$; $r_M<r<r_\mr{abs}$. In the GRANIT experiment these screens could correspond to the positions of the neutron mirror and the neutron absorber, respectively. We will describe the system in terms of the screen $\cS^{\supset m}_{r_\mr{abs}}$ that is known to contain the particle $m$. The entropy of this screen depends on the total energy $E$ of the screen and on the macroscopic parameter $\vec{r}$ that describes the position of the particle $m$:
\be\label{S.screen}
S(E,\vec{r}) = \log\Omega(E,\vec{r}) \,,
\ee
where $\Omega(E,\vec{r})$ is the number of the microstates on $\cS^{\supset m}_{r_\mr{abs}}$ that are associated with the mass distribution: $M$ at origin and $m$ at $\vec{r}$. In EG, it is postulated that the number of microstates $\Omega(E,\vec{r})$ increases when the parameter $\vec{r}$ is infinitesimally changed to $\vec{r}+d\vec{r}$ with $d\vec{r}\cdot\vec{r}<0$. Hence the number of microstates in the macrostate of $\cS^{\supset m}_{r_\mr{abs}}$ increases when the particle $m$ is displaced closer to the mass $M$. Now the crucial question reads, does this increase in the number of microstates $\Omega(E,\vec{r})$ imply that the number of states in the density operator of the particle $m$ increases too? If we assume that the density operator of the particle at $\vec{r}$ consists of fragments of the $\Omega(E,\vec{r})$ microstates on $\cS^{\supset m}_{r_\mr{abs}}$ and the number of linearly independent fragments is proportional to $\Omega(E,\vec{r})$, it is clear that the number of states in the density operator of the particle $m$ does depend on the position $\vec{r}$, and thus the general idea of the argument in Refs.~\cite{KobakhidzeEG,KobakhidzeEG2} would be correct.  A state of the particle $m$ would inevitably lose its coherence if it extends in the direction of gravity, because the translation operator for the state of the particle would be nonunitary in this direction. We can now see that the assumption
\begin{gather}
\text{\textit{``the density operator of a particle within a holographic screen}}\nn\\
\text{\textit{consists of fragments of the microstates of the screen}}\label{fragmentpremise}\\
\text{\textit{and the number of linearly independent fragments is}}\nn\\
\text{\textit{proportional to the number of microstates''}}\nn
\end{gather}
is the decisive premise behind the analysis of Refs.~\cite{KobakhidzeEG,KobakhidzeEG2}. 
The latter part of this premise means that the number of linearly independent fragments in the density operator of the particle is assumed to be a monotonically increasing function of the number of microstates $\Omega(E,\vec{r})$. Such an assumption is necessary in order to draw a conclusion on the $\vec{r}$-dependence of the entropy of the particle.
For example, if the density operator of the particle consists of $[\Omega(E,\vec{r})]^{m/E}$ independent fragments, and each of those states is equally probable, the entropy of the particle would be the fraction $m/E$ of the entropy of the screen \eqref{S.screen}.
Although the assumption \eqref{fragmentpremise} may seem to be reasonable, there are reasons to suspect that it may be invalid.

Let us first discuss what happens when we abandon the assumption that the number of linearly independent fragments is proportional to $\Omega(E,\vec{r})$. Indeed, we do not see any compelling reason to accept this assumption.
A certain fragment of two linearly independent microstates can be linearly dependent states. Such dependent fragments are in fact identical if the microstates in the density operator of the screen are orthonormal as is the case in spectral representation.
In the density operator of the particle such identical fragments count as one state with double the probability, rather than two different states. Let us generalize this point. The fragments of the $\Omega(E,\vec{r})$ microstates that are assumed to describe the particle can be divided into $K$ sets labelled by $i=1,2,\ldots,K \le \Omega$, which each consist of $k_i$ identical states; $\Omega=\sum_{i=1}^K k_i$, $1\le k_i\le\Omega$.
Both $K$ and $k_i$ may depend on the mass distribution and on the particle, in particular its position $\vec{r}$. Now the density operator of the particle consists of $K(E,\vec{r})$ states which have the probabilities $p_i=k_i(E,\vec{r})/\Omega(E,\vec{r})$, assuming the $\Omega(E,\vec{r})$ microstates of the screen form a microcanonical ensemble with equal probability $1/\Omega(E,\vec{r})$ for each microstate. Thus the particle has the entropy
\be\label{S.particle}
S_m(E,\vec{r}) = -\sum_{i=1}^{K(E,\vec{r})} \frac{k_i(E,\vec{r})}{\Omega(E,\vec{r})} \log\frac{k_i(E,\vec{r})}{\Omega(E,\vec{r})} \,,
\ee
which is smaller than the entropy of the screen \eqref{S.screen},
unless all the fragments of the microstates are unique, i.e., $k_i=1$ and $K=\Omega$. The smaller the particle $m$ is compared to the screen $E$, the fewer sets $K$ of fragments that contain higher numbers $k_i$ of identical states and consequently the smaller amount of entropy \eqref{S.particle} can be expected. Since we do not know how $K$ and $k_i$ depend on $\vec{r}$, there is currently no way of telling how \eqref{S.particle} behaves when the position $\vec{r}$ of the particle changes. Conceivably there could even exists a microscopic theory where the entropy \eqref{S.particle} of a particle is independent of the position of the particle, although the entropy \eqref{S.screen} of the screen necessarily depends on $\vec{r}$. That would require that $K$ and $k_i$ depend on $\vec{r}$ in a way that cancels the $\vec{r}$-dependence of $\Omega$, e.g., so that $K$ and $p_i$ are independent of $\vec{r}$. Therefore we can conclude that by only assuming that a particle is described by fragments of the microstates of a screen one cannot draw any definitive conclusion on the $\vec{r}$-dependence of the entropy of the particle. Such a conclusion would require some knowledge on the $\vec{r}$-dependence of $K$ and $k_i$, which we do not possess.

The first reasons that lead us to doubt the validity of the assumption \eqref{fragmentpremise} are the inconsistencies found in the treatment of neutron states in Refs.~\cite{KobakhidzeEG,KobakhidzeEG2}. We have discussed these inconsistencies in the previous sections at some length. They were originally pointed out in Ref.~\cite{ourEGnote} albeit briefly. These inconsistencies in Refs.~\cite{KobakhidzeEG,KobakhidzeEG2} are caused by a misinterpretation concerning the microstates of holographic screens. It should be possible to correct these inconsistencies, e.g. by using the more global point of view presented in this section, without jeopardizing the main idea of the argument and the result.

The last reason to question the assumption \eqref{fragmentpremise} is more conceptual. According to the principle of equipartition of energy over the microscopic degrees of freedom on a holographic screen, it is possible to estimate the number of microscopic degrees of freedom that are associated with a certain particle within the screen. However, in the absence of a microscopic theory, we do not know how the state of a particle within a screen is encoded in the microscopic degrees of freedom on the screen. We are not convinced that it is possible to dissect a microstate of a holographic screen into separate parts which describe the states of particles at different positions within the screen. This is essentially what the assumption \eqref{fragmentpremise} claims to be possible. The microstates of a holographic screen are very nonlocal states -- this is essential to the idea of holography. Until we know some details on the nature of the microstates of holographic screens, we should not try to dissect them into fragments.

The assumption \eqref{fragmentpremise} would actually have even more far reaching consequences than we noted above. It would imply that the size of the state space of any particle depends on the position parameter of every particle. This would mean that if even one of the particles within a holographic screen can be displaced in the emergent direction, the coherence of the states of every particle inside the screen would be destroyed. In fact, a similar reasoning could be applied to any theory of holographic emergent gravity where the entropy of a holographic screen depends on the distribution of matter it contains. We can generalize this reasoning even further by replacing the concept ``holographic screen'' with any microscopic system that consists of a finite number of degrees of freedom. Consider a generic theory of emergent space that is based on such a finite microscopic system. If the entropy of the microscopic system depends on the distribution of matter in the emergent space, the generalization of the postulate \eqref{fragmentpremise}
\begin{gather}
\text{\textit{``the density operator of a particle in the emergent space consists of}}\nn\\
\text{\textit{fragments of the microstates of the microscopic system}}\label{fragmentpremise2}\\
\text{\textit{and the number of linearly independent fragments is}}\nn\\
\text{\textit{proportional to the number of microstates''}}\nn
\end{gather}
implies that the theory is unable to support coherent quantum states in the emergent space. We have clearly learned something general and interesting on the nature of microstates in theories of emergent space. Either the entropy of the microscopic system cannot depend on the distribution of matter in the emergent space, which would mean that entropic forces of any kind cannot exist in the emergent space, or the assumption \eqref{fragmentpremise2} is incorrect.

Lastly, we discuss the emergence of the state of a particle in EG. Since we know that the screen $\cS^{\supset m}_{r_\mr{abs}}$ contains the particle $m$ and the screen $\cS_{r_M}$ does not contain the particle, we should be able to obtain the state of the particle $m$ by coarse graining the screen $\cS^{\supset m}_{r_\mr{abs}}$ down to the screen $\cS_{r_M}$. In this process the macrostate $\rho_{\cS^{\supset m}_{r_\mr{abs}}}(\vec{r})$ of the screen $\cS^{\supset m}_{r_\mr{abs}}$ reduces to the product of the macrostate $\rho_{\cS_{r_M}}(\vec{r})$ of the screen $\cS_{r_M}$ and the state of the particle $\psi_m(\vec{r})$ between the screens at $\vec{r}$:\footnote{The state $\rho_{\cS_{r_M}}(\vec{r})$ of the screen $\cS_{r_M}$ is not strictly independent of the particle $m$ at $\vec{r}$ even though the screen does not contain the particle, because in EG the particle is postulated to affect the information on the screen from a distance.}
\be
\rho_{\cS^{\supset m}_{r_\mr{abs}}}(\vec{r}) \ \xrightarrow{\ \text{coarse graining}\ }\ \rho_{\cS_{r_M}}(\vec{r}) \otimes \psi_m(\vec{r}).
\ee
The coarse graining process gets rid of all the unnecessary microscopic information and gives rise to the macroscopic state of the matter in the emerged space. The state of the particle $\psi_m(\vec{r})$ would have all the relevant information built into it, such as boundary conditions and gravity. The state of the particle $\psi_m(\vec{r})$ in the emerged space can be a very different kind of state compared to the microstates of the screens. The fact that the number of microstates $\Omega(E,\vec{r})$ in $\rho_{\cS^{\supset m}_{r_\mr{abs}}}(\vec{r})$ depends on $\vec{r}$ does not necessarily mean that the number of states in $\psi_m(\vec{r})$ depends on $\vec{r}$ in a similar way. The state space of the particle in emerged space does not need to depend on the position of the particle. Thus the translation operator for the state $\psi_m(\vec{r})$ could be unitary. Hence coherence and interference could exist in such description. It is conceivable that such a description could realize all the known quantum mechanical phenomena. Presumably, in order to derive the state of the particle $m$ from EG in this way, we need some knowledge on the microscopic theory of holographic screens, including details of the coarse graining procedure. Unfortunately, we do not possess this knowledge yet.

\section{Conclusion}\label{sec.6}
It is of course possible and even expectable that sooner or later EG \cite{Verlinde} will be refuted on some experimental or theoretical grounds. Phenomena that involve quantum effects such as gravitationally-bound quantum states of neutrons \cite{N} and gravitationally induced quantum interference \cite{Colella} are certainly good candidates for the search of such proof.
However, our arguments show that the treatment of EG in Refs. \cite{KobakhidzeEG,KobakhidzeEG2} contains a misinterpretation concerning the relation of the microstates of holographic screens and the state of a particle in the emergent space. It leads to at least two inconsistencies. A point of view that could resolve the inconsistencies was presented in Sect.~\ref{sec.5}. We identified the decisive premise \eqref{fragmentpremise} behind the analysis of Refs.~\cite{KobakhidzeEG,KobakhidzeEG2} and presented arguments that suggest this assumption may be incorrect. Thus it should not be accepted without proper justification.
Therefore the analysis in Refs. \cite{KobakhidzeEG,KobakhidzeEG2} cannot quite yet be accepted as proof that EG fails to explain the observation of gravitationally-bound quantum states of neutrons and the experiments that demonstrate gravitationally induced quantum interference.

In summary, the fate of EG depends on the validity of the assumption \eqref{fragmentpremise}. If the assumption \eqref{fragmentpremise} is correct in the holographic description assumed in EG, then EG \cite{Verlinde} fails to explain the results of the experiments \cite{N,Colella}. We suspect that more detailed knowledge on EG is required in order to pass a conclusive judgement on this matter.

In Sect.~\ref{sec.5} we showed that if the state of a particle in emergent space can be considered to consists of fragments of the microstates of the microscopic theory, that will not only imply that EG is unable to reproduce standard quantum mechanics in the presence of gravity. It would mean that any theory of emergent space, where the entropy of the underlying microscopic system depends on the distribution of matter in the emergent space, would in general be unable to describe coherent quantum states in the emergent space. Thus, if the generalization \eqref{fragmentpremise2} of the premise \eqref{fragmentpremise} is valid, any such theory could be refuted by experiments such as \cite{N,Colella}.

Unfortunately, we do not know the microscopic theory of holographic screens and therefore we cannot fully analyze the mentioned neutron experiments from the point of view of a screen that contains the hole system that is relevant to the state of a test particle. It is conceivable that such a description could realize all the known quantum mechanical phenomena. Due to the lack of a microscopic theory we are still bound to consider a situation where the particle is in the emergent space above a screen. In the emergent space the particle should be described in terms of a theory defined in actual space, e.g. quantum mechanics or quantum field theory, not in terms of microscopic degrees of freedom of a screen. Naturally, if a microscopic theory of holographic screens is unable to reproduce the predictions of such well-established theories, it must be rejected. Since the microscopic origin of EG is still very vague, it is not easy to prove such failure.

We do not actually expect that physics in emergent space could be precisely that of standard quantum mechanics or quantum field theory. When the microscopic system behind space contains a finite amount of information, emergent space cannot be continuous but rather discrete or quantized in some way. Hence a more precise description of physics in emergent space at short distances should be given in some framework with a minimal distance, such as for example quantum mechanics and quantum field theory on noncommutative spaces.

We expect that the notion of graviton will become necessary for the understanding of gravitational phenomena \cite{ourEGnote}. Therefore, for EG to become a viable theory of gravity, it will have to be able to accommodate graviton as an emergent concept, much like that in AdS/CFT duality or as phonon in solid state physics.
The idea of classicalization of gravitons \cite{D1,D2,D3,D4} could provided an interesting point of view to EG. In the classicalization phenomenon, an interaction with a center-of-mass energy higher than the Planck mass produces a classical black hole configuration -- a classicalon -- which may self-unitarize gravity at high energies. The entropy of the black hole has a precursor given by the number of soft quanta composing the classicalon, and it could even be described in terms of classicalon states.
This could provide a way to gain new understanding on the nature of entropy and microstates in EG.

\paragraph{Acknowledgements}
We are indebted to Gia Dvali for illuminating discussions and wish to thank Valery Nesvizhevsky for discussions on the GRANIT experiment. 
We are grateful to Archil Kobakhidze for critical comments and correspondence and for providing us with his manuscript \cite{KobakhidzeEG2} before its submission to the arXiv. 
M.O. is supported by the Jenny and Antti Wihuri Foundation. 
The support of the Academy of Finland under the Project Nos. 136539 and 140886 is gratefully acknowledged.

\end{document}